\documentclass[preprint,prc,aps,superscriptaddress,nofootinbib,showpacs]{revtex4}
\usepackage[dvips]{graphicx}
\usepackage{amsfonts, color, float, bbm, amsmath}
\usepackage{amssymb}

\def\lp {\left( }
\def\rp {\right) }
\def\lb {\left[ }
\def\rb {\right] }
\def\lc {\left\{ }
\def\rc {\right\} }
\def\ra {\rangle }
\def\la {\langle }
\def\nn {\nonumber}

\def\beq{\begin{equation}}
\def\eeq{\end{equation}}
\def\bea{\begin{eqnarray}}
\def\eea{\end{eqnarray}}
\def\ni{\noindent}

\def\d {\partial }

\def\cd {\!\cdot\!}

\def\rar {\rightarrow}

\def\ub {\bar u}
\def\pb {\bar p}

\def\Ps {\not\!P}
\def\qs {\not\!q}
\def\Qs {\not\!\!Q}

\def\st {\tilde{\s}}

\def\Pb {\bar{\Pi}}

\def\Sb {\bar{S}}

\def\sp {\!+\!}
\def\sm {\!-\!}

\def\vs {\vspace}

\def\D {\Delta}

\def\g{\gamma}
\def\D {\Delta}
\def\G {\Gamma}

\def\l {\lambda}

\def\m{\mu}
\def\n{\nu}
\def\o{\omega}

\def\p{\pi}
\def\P{\Pi}
\def\r{\rho}
\def\s{\sigma}

\def\cL {{\cal L}}
\def\cO {{\cal O}}

\def\bfi {\mbox{\boldmath $\phi$}}
\def\bk {\mbox{\boldmath $k$}}
\def\br {\mbox{\boldmath $r$}}

\def\bx {\mbox{\boldmath $x$}}

\def\bnb {\mbox{\boldmath $\nabla$}}

\def\bp {\mbox{\boldmath $p$}}

\def\bpi {\mbox{\boldmath $\pi$}}
\def\btau {\mbox{\boldmath $\tau$}}

\def\bq {\mbox{\boldmath $q$}}

\begin{document}

\title{Pion-delta sigma-term}


\author{I. P. Cavalcante}
\email[]{ipcavalcante@dfi.ufms.br}
\affiliation{Depto. de F\'{\i}sica, CCET, Universidade Federal de
Mato Grosso do Sul,\\
C.P. 549, C.E.P. 79070-900, Campo Grande, MS, Brazil.}

\author{M. R. Robilotta}
\email[]{robilotta@if.usp.br}
\affiliation{Instituto de F\'{\i}sica, Universidade de S\~{a}o Paulo,\\
C.P. 66318, 05315-970, S\~{a}o Paulo, SP, Brazil.}

\author{J. S\'a Borges}
\email[]{saborges@uerj.br}
\affiliation{Universidade do Estado do Rio de Janeiro, Instituto de
F\'{\i}sica,\\
Rua S\~ao Francisco Xavier, 524, Maracan\~a, Rio de Janeiro, RJ, Brazil.}

\author{D. de O. Santos}
\affiliation{Depto. de F\'{\i}sica, CCET, Universidade Federal de
Mato Grosso do Sul,\\
C.P. 549, C.E.P. 79070-900, Campo Grande, MS, Brazil.}

\author{G. R. S. Zarnauskas}
\email[]{gabrielz@if.usp.br}
\affiliation{Instituto de F\'{\i}sica, Universidade de S\~{a}o Paulo,\\
C.P. 66318, 05315-970, S\~{a}o Paulo, SP, Brazil.}


\begin{abstract}We use a configuration space chiral model in order to
evaluate nucleon and delta $\s-$terms. Analytic expressions are
consistent with chiral counting rules and give rise to expected
non-analytic terms in the chiral limit. We obtain the results
$\s_N=46$ MeV and $\s_\D=32$ MeV, which are very close to values
extracted from experiment and produced by other groups.  
\end{abstract}


\maketitle

\vspace{5mm}

\section{introduction}

The delta ($\D$) plays a very important role in low-energy
pion-nucleon ($\p N$) scattering and correlated processes, such as
the nucleon-nucleon interaction. Its contribution as an intermediate
state, in many instances, supersedes that of the the nucleon. This
happens for two main reasons. The first one is that the $\p N\D$
coupling constant is rather large, whereas the other is related to
chiral symmetry. 

At low energies, pion-hadron  interactions are well described by
effective theories, in which an approximate $SU(2)\times SU(2)$
symmetry, broken by the pion mass $(\m)$, accounts for the smallness
of the $u$ and $d$ quark masses. In this framework, elastic
pion-baryon scattering is dominated by diagrams involving both contact
terms and propagating states. In order to comply with threshold chiral
theorems, the latter are typically given by polynomials in small
quantities, such as the pion mass or three-momenta, divided by energy
denominators. When the delta is present, the scale of some
denominators is given by the quantity $\o_\D=(M^2-m^2-\m^2)/2m$, where
$M$ and $m$ are respectively the delta and nucleon masses. As the
difference $\D\equiv M-m$ is small, one has $\o_\D\sim \D$. Delta
contributions are given by ratios of small quantities and may turn out
to be large. In such cases, numerical values adopted for $\D$ do
influence predictions produced by effective theories, especially those
that rely on the small scale expansion\cite{HHK} or the heavy baryon
approximation\cite{HB}. 

In chiral perturbation theory, there is a clear conceptual distinction
between the bare baryon masses, present in the lagrangian, and their
observed values, which include loop corrections. The former should, in
principle, be preferred as inputs in the evaluation of theoretical
amplitudes. Nevertheless, as there is little knowledge available
concerning the bare delta mass, one tends to use physical values in
calculations. In most cases, it is reasonable to expect that this
would have little numerical importance. On the other hand, in the case
of the parameter $\D$, which is a small quantity, the influence of
loops may become relatively large. 

Recently Bernard, Hemmert and Meissner\cite{BHM}, have stressed that
the value of $\D_0$, the delta-nucleon mass splitting in the chiral
limit, is an important constraint to lattice data
extrapolation. The purpose of the present work is to estimate the
delta $\s$-term, which controls the change induced in $\D$ when one
goes from bare to physical masses. This $\s$-term was studied  in the
framework of a quark model by Lyubovitskij, Gutsche, Faessler and
Drukarev\cite{LGFD} and the reader is referred to their paper for a
clear formulation of the problem and earlier works.  

According to the Feynman-Hellmann theorem\cite{FH} the mass $m_B$ of a
baryon $B$ is related to its sigma-term $\s_B$ by $\s_B=\m^2\, d\,
m_B/ d\m^2$. Therefore the sigma-term provides a measure of the shift
in the baryon mass due to chiral symmetry breaking. Whenever it is
possible to evaluate $\s_B$ as a function of $\m$, the bare mass
$m_{B_0}$ can be extracted from the relation  
\beq
m_B = m_{B_0} + \int_0^{\m^2} d\l \; \s_B(\l) / \l \;.
\label{1.1}
\eeq

As the leading term in $\s_B$ is proportional to $\m^2$, it enters
directly the mass shift and the difference $m_B\!-\!\s_B$ already
provides a crude estimate for the bare value. In the case of the
nucleon, one has $\s_N$=45 MeV\cite{GLS}, which amounts to 5\% of its
physical mass. In chiral perturbation theory, the leading contribution
to $\s_N$ cannot be predicted theoretically. Formally, it is associated
with the constant $c_1$ of the second order lagrangian\cite{GSS,BL},
which can be extracted from empirical subthreshold information. The
situation of the delta is much worse, for $\p\D$ scattering data are
not available. One is then forced to resort to models in order to
calculate the delta $\s$-term, which is associated with the parameter
$a_1$ defined in ref.\cite{HHK}. 

In this work we estimate $\s_\D$ using a model which proved to be
successful in the case of the nucleon. Our paper is organized as
follows. In section \ref{2} we review our calculational procedure in the
case of the nucleon and present results for the delta in section \ref{3},
leaving technical details to the appendices. The main expressions for
both the nucleon and delta $\s$-terms in configuration space are given
in appendix \ref{B}, written in terms of the loop integrals defined in
appendix \ref{A}. The consistency of our results with standard chiral
counting rules is discussed in appendix \ref{C} whereas their behaviour in
the chiral limit is given in appendix \ref{D}. A summary is provided in
section \ref{4}.

\section{model for the sigma term}
\label{2}

\begin{figure}[H]
\begin{center}
\includegraphics[width=0.70\columnwidth,angle=0]{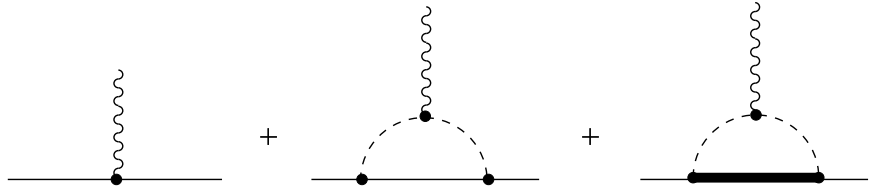}
\caption{Contact term and triangle diagrams contributing to the $\s$-term.} 
\label{fig:diagram}
\end{center}      
\end{figure}

In order to evaluate $\s_\D$, we follow a procedure used previously in
the study of $\s_N(t)$, the nucleon scalar form factor\cite{R}, which
is briefly reviewed here. The leading contributions to this function
is $\cO(q^2)$ whereas the triangle diagram, involving only known
masses and coupling constants, gives rise to corrections which begin
at $\cO(q^3)$ and are completely determined. At $\cO(q^4)$, on the
other hand, interactions incorporate the low energy constants $c_1,
c_2$ and $c_3$. Data on $\p N$ subthreshold coefficients indicate that
$c_2$ and $c_3$ are larger than $c_1$ and that their values are
approximately saturated by $\D$ intermendiate states\cite{BL}. Thus,
up to $\cO(q^4)$, the function $\s_N(t)$ can be well represented by
the leading tree contribution associated with $c_1$, supplemented by
the two triangle diagrams shown in fig.\ref{fig:diagram}, involving
$N$ and $\D$ intermediate states. In the sequence we will make use of
the fact that, in configuration space, contact and loop contributions
split apart, since the Fourier transform acts as a filter\cite{L}. As
a result, the theoretically undetermined leading tree term yields a
zero-range $\delta$-function, whereas the triangle diagrams give rise
to spatially distributed structures, fully determined by known
parameters.

\vspace{3mm}
\ni

\ni

\vspace{3mm}

The nucleon scalar form factor in momentum space is defined by 
\beq
\la N(p') | \sm \cL_{sb}\, | N(p) \ra = \s_N(t) \; \ub(p')\; u(p) \;,
\label{2.1}
\eeq

\ni
where $\cL_{sb}$ is the symmetry breaking term in the lagrangian and
$t=(p' \sm p)^2$. In terms of the quark degrees of freedom, one has
$\cL_{sb} = - \hat{m}\,(\bar{u}u \sp \bar{d}d)$, with $\hat{m}=(m_u+
m_d)/2$. The configuration space scalar form factor is denoted by
$\tilde{\s}_N$ and given by  
\beq
\tilde{\s}_N(\br)=  \int \frac{d^3q}{(2\p)^3} \; e^{-i \,\bq \cdot \br}\; \s_N(t)
\label{2.2}
\eeq 

\ni
with $\bq = (\bp' \sm \bp)$, in the Breit frame.
The nucleon $\s$-term, defined as $\s_N\equiv\s_N(t=0)$, is given by 
\beq
\s_N = 4\p \int_0^\infty dr\;r^2\;\st_N(r)\;.
\label{2.3}
\eeq

The contributions from the diagrams of fig.\ref{fig:diagram} to $\tilde{\s}_N(\br)$ read 
\beq
\tilde{\s}_N(\br) = - 4\, c_1\, \m^2\, \delta^3(\br) +
\tilde{\s}_{N_N}(r) + \tilde{\s}_{N_\D}(r) \;, 
\label{2.4}
\eeq

\ni 
where $\tilde{\s}_{N_N}(r)$ and $\tilde{\s}_{N_\D}(r)$ are given by
eqs.(\ref{b.4}) and (\ref{b.5}) of  appendix \ref{B} and displayed in
fig.\ref{fig:snucleon}. These functions are based on unregularized loop integrals and
diverge for small values of $r$. In momentum space, regularization is
achieved by means of counterterms, which give rise to polynomials in
$t$, designed to cancel the divergences of the loop integrals. In
configuration space, this regularization procedure amounts to adding
$\delta$-functions and their derivatives to $\tilde{\s}_N(r)$. This
gives rise to a regularized form factor which is very large both at
$r=0$ and in a sizeable vicinity of that point. We argue, in the
sequence, that this picture is not consistent with the definition of
the form factor given by eq.(\ref{2.1}). 

\vspace{3mm}
\ni

\begin{figure}[h]
\begin{center}
\includegraphics[width=0.7\columnwidth,angle=-90]{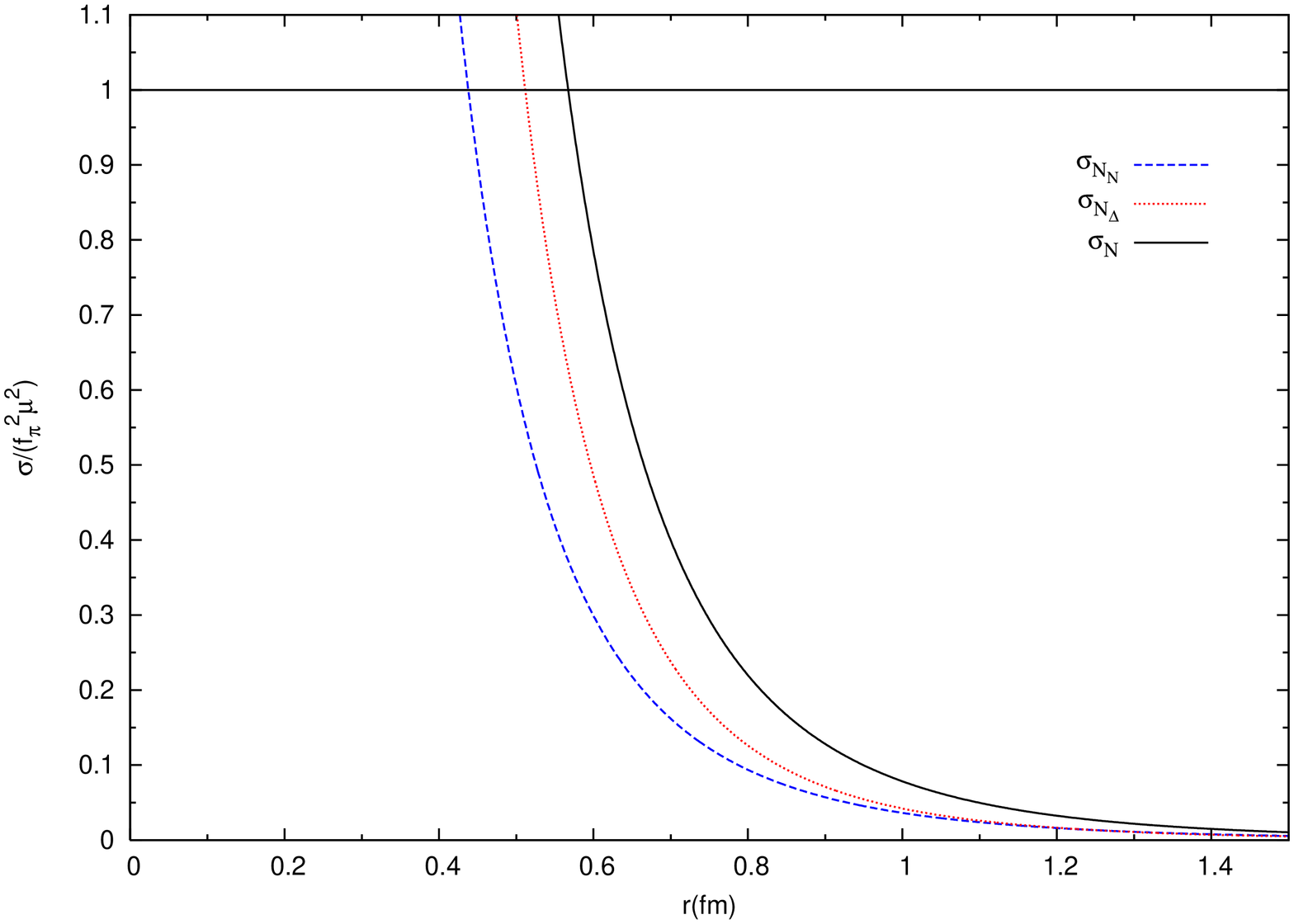}
\caption{Spatial dependence of the nucleon scalar form factor
(continuous line) and partial contributions due to $N$ and $\D$
intermediate states  [eq.(\ref{b.4}), dashed line and eq.(\ref{b.5}),
dotted line, respectively].}  
\label{fig:snucleon}
\end{center}      
\end{figure}

\ni

\vspace{3mm}

Pions are Goldstone bosons, collective states derived from the
$q\bar{q}$ condensate. The corresponding degrees of freedom are
appropriately accomodated  into non-linear lagrangians and described
by the field $U=\exp(i\btau\cdot\hat{\bpi}\;\theta)$, where
$\hat{\bpi}$ is the isospin direction and $\theta$ is the chiral
angle. This function can be expressed as $U=\cos\theta + i \btau\cdot
\hat{\bpi}\, \sin\theta$ and the dimensional pion field is given by  
\beq
\bfi = f_\p \sin\theta \,\hat{\bpi}\;.
\label{2.5}
\eeq

Long ago, Skyrme\cite{Sky}, in a series of papers, considered the
possibility of pion fields being either {\em weak} or {\em
strong}. It is worth noting that the words {\em weak} and
{\em strong}, as used here, are akin to the notion of weak and strong
electromagnetic fields developed by Schwinger, and {\bf not} at
all related to the nature of the fundamental interactions. In the
former case, changes in  the $q\bar{q}$ condensate are small, one
relies on the approximation $ \bfi \simeq f_\p \theta \,\hat{\bpi}$
and can employ perturbative techniques, as in chiral perturbation
theory (ChPT). In the latter, disturbances of the QCD vacuum become
important and the non-linear nature of pionic interactions manifests
itself through the condition $|\bfi| \leq f_\p$. The physical picture
behind eq.(\ref{2.5}) is that pions, as Goldstone bosons, destroy the
$q\bar{q}$ condensate in order to exist. 

When strong fields are present, constraints also apply to the scalar
form factor. The symmetry breaking lagrangian is written in terms of
the dimensional pion field as 
\beq
\cL_{sb} = \frac{1}{4} f_\p^2 \, \m^2 \, Tr \lb U+ U^\dagger\rb  
= f_\p^2 \, \m^2 \,\cos\theta \;.
\label{2.6}
\eeq

This structure shows that $\cL_{sb}$ is a bound function and
definition (\ref{2.1}) means that the same necessarily happens with
the scalar form factor. The function $\tilde{\s}_N(x)$ corresponds to
a mass density induced in the vacuum by the presence of the nucleon,
which manifests itself in the form of a pion cloud. Far away from the
nucleon, eq.(\ref{2.6}) yields the density of the condensate, which is
negative and equal to $-f_\p^2\,\m^2$. In the description of a
nucleon, it is convenient to use a convention for the energy in which
the density tends to zero at long distances and $\cL_{sb}$ is
rewritten as  
\beq
\cL_{sb} = f_\p^2 \, \m^2 \,(\cos\theta-1) \;.
\label{2.7}
\eeq 

In this new convention, the density vanishes when $r\rar \infty$ and
increases monotonically as one approaches the center of the nucleon as
in fig.\ref{fig:snucleon}. At  a critical radius $R$ one has $\cos\theta=1$, the
density becomes that of empty space and the condition  
\beq
\tilde{\s}_N(R) = f_\p^2\,\m^2
\label{2.8}
\eeq 

\ni 
holds.
Beyond this point, a further increase in $\tilde{\s}_N$ would
correspond to $\cos\theta>1$. In order to prevent this behaviour, we
assume that the condensate no longer exists in the region $r<R$, and
that the energy density saturates at $r=R$. For this reason, in our
previous evaluation of $\s_N$\cite{R},  we used the expression 
\beq
\s_N = \frac{4}{3} \p R^3 \;f_\p^2 \m^2+ 4\p \int_{R}^\infty dr\;r^2\;\st_N(r)
\label{2.9}
\eeq

\ni 
instead of eq.(\ref{2.3}).
This procedure is the basis of our model.

In the numerical determination of $\s_N$, we use the results of
appendix \ref{B} and consider two possibilities for the $\p N \D$ coupling
constant in the lagrangian (\ref{b.2}), corresponding to either
the $SU(4)$ prediction $g_{\p N\D}= 3 \,g_A / 2\sqrt{2}= 1.33$ or
$g_{\p N \D}= 1.47$, which yields $\G$=120MeV for the $\D$ decay
width. The corresponding results, given in table \ref{T1}, are quite close to
the value extracted from experiment by Gasser, Leutwyler and
Sainio\cite{GLS}, namely $\s_N= 45$MeV. 


\ni

\begin{table}[h]
\begin{center}
\caption{Nucleon $\s$-term as function of the $\p N \D$ coupling constant.}
\vspace{3mm}
\begin{tabular} {|c|c|c|}
\hline 
$g_{\p N\D}$ &  $R$ (fm) & $\s_N$ (MeV) 	\\ \hline
\hline
1.33         & 0.57      & 45.8        		\\ \hline 
1.47  	     & 0.59	 & 49.4        		\\ \hline 

\end{tabular}
\label{T1}
\end{center}
\end{table}


Consistency with chiral symmetry is an important issue in this problem.
Therefore, we note that, although the chiral powers of the pion mass
expected from triangle diagrams are not explicit in the expressions of
appendix \ref{B}, the use of covariant relations among integrals\cite{HR}
allows results for partial contributions to the $\s$-term to be recast
in such a way that these powers become apparent, as shown in appendix
\ref{C}. In appendix \ref{D} we show that the formal chiral expansion of
eq.(\ref{2.9}) gives rise to the expected non-analytic terms ($\log
\m$ and $\m^3$) and agrees fully with that produced by standard chiral
perturbation theory\cite{BL}, provided the renormalization scale is
identified with $1/R$.

\section{delta $\s$-term}
\label{3}

The delta scalar form factor is defined as
\beq 
<\D (p',s') | - {\cal L}_{sb} | \D(p,s) > \equiv - \ub_\m^{s'} (\bp')
\lb g^{\m\n} \s_\D (t) + p'^\n p^\m F_T (t) \rb  \;u_\n^s (\bp) \; , 
\label{3.1}
\eeq

\ni
where $ u_\n^s $  is the $\D$ spinor\cite{D} and $\s_\D$ and $F_T$ are
respectively the scalar and tensor form factors. The minus sign on the
r.h.s. is associated with the conventions used in the free $\D$
lagrangian as in ref.\cite{HHK}. We assume that the scalar form factor
is determined by a short range contact interaction and the two long
range two-pion processes shown in fig.\ref{fig:diagram}. 

\begin{figure}[h]
\begin{center}
\includegraphics[width=0.7\columnwidth,angle=-90]{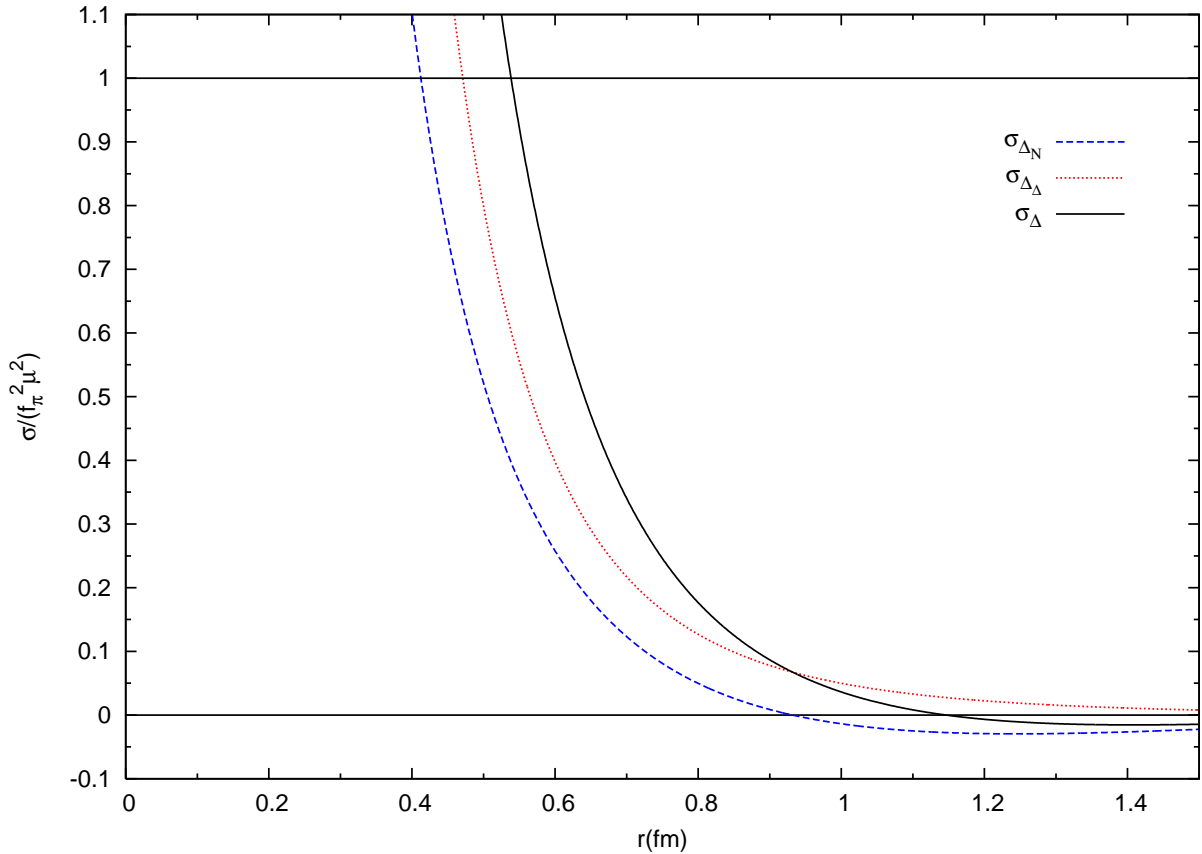}
\caption{Spatial dependence of the delta scalar form factor
(continuous line) and partial contributions due to $N$ and $\D$
intermediate states  [eq.(\ref{b.9}), dashed line and eq.(\ref{b.10}),
dotted line, respectively].}  
\label{fig:sdelta1}
\end{center}      
\end{figure}

\begin{figure}[H]
\begin{center}
\includegraphics[width=0.6\columnwidth,angle=-90]{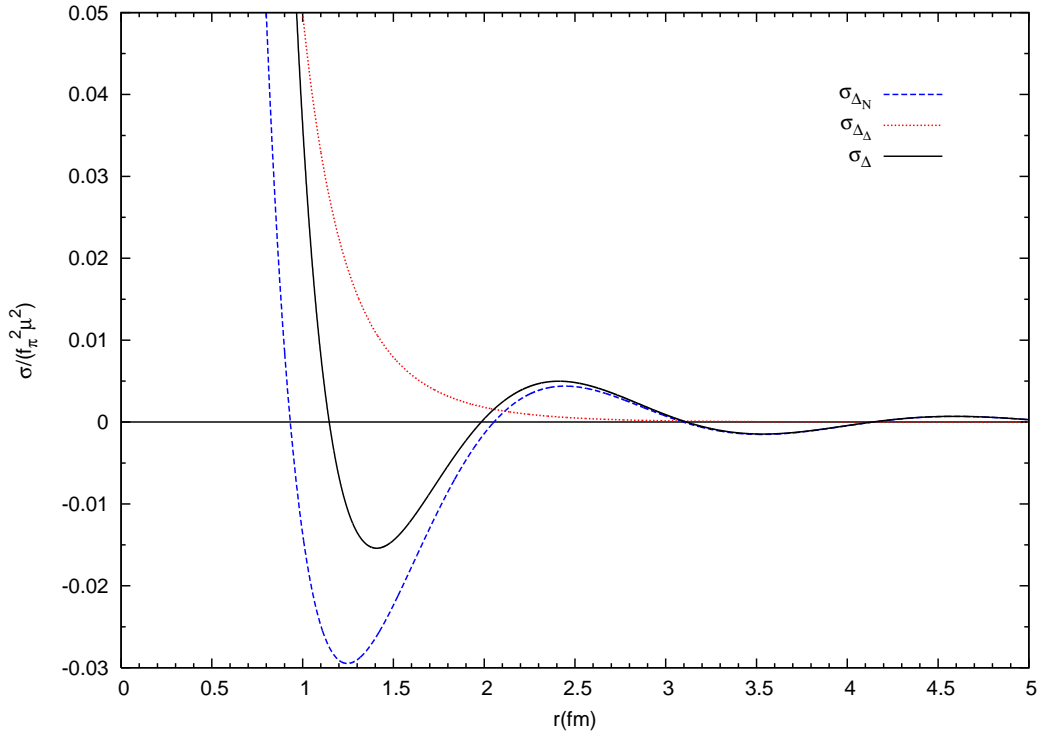}
\caption{Expanded portion of fig.\ref{fig:sdelta1}, with the same
conventions.}   
\label{fig:sdelta2}
\end{center}      
\end{figure}

In figs. \ref{fig:sdelta1} and \ref{fig:sdelta2} (zoom in) we display the
profile functions for the partial contributions to $\s_\D$ given by
eqs.(\ref{b.9},\ref{b.10}) and it is interesting to note that the
nucleon contribution oscillates in the outer region, in sharp contrast
with fig.\ref{fig:snucleon}. This behaviour is due to
the fact that the delta is unstable and makes $\s_\D$ to be smaller
than $\s_N$. The structure of partial contributions for $SU(4)$
coupling constants is given in table \ref{T2}, where core and cloud refer
respectively to regions inside and outside the cutting radius $R$.

\vspace{3mm}
\ni

\vs{3mm}

\ni
\vspace{3mm}

\ni
\vspace{3mm}

\ni

\begin{table}[h]
\begin{center}
\caption{Partial contributions to $\s_N$ [eqs.(\ref{b.4},\ref{b.5})]
and $\s_\D$ [eqs.(\ref{b.9},\ref{b.10})].} 
\vspace{3mm}
\begin{tabular} {|c|c|c|c|c|}
\hline
		& core	& cloud $N$ & cloud $\D$ & sum 	\\ \hline
\hline
$\s_N$ (MeV)	& 16.7	& 13.0	    & 16.1	 & 45.8	\\ \hline
$\s_\D$ (MeV)	& 14.3	& -1.5	    & 19.3	 & 32.1	\\ \hline

\end{tabular}
\label{T2}
\end{center}
\end{table}

\vspace{3mm}

Processes containing nucleon intermediate states give rise to an
imaginary component $\s_\D^I$ for the delta $\s$-term, which can be
related to the decay width by means of the Feynman-Hellmann
theorem\cite{FH}: 
\beq
\s_\D - i\, \s_\D^I =  \m^2\; \frac{d\, (M - i\, \G/2) }{d\m^2}.
\label{3.2}
\eeq

Using\cite{BL}
\bea
&&\G= \frac{g_{\p N \D}^2}{24 \p M^2 f_\p^2} \, q_\D^3 \lb (M \sp m)^2 -\m^2 \rb \;, 
\nn\\[2mm]
&& q_\D= \frac{1}{2M}\,\sqrt{M^4 \sp m^4 \sp \m^4 \sm 2 m^2 M^2 \sm 2
\m^2 M^2 \sm 2 \m^2 m^2} \;, 
\label{3.3}
\eea

\ni
one finds 
\beq
\s_\D^I = - \frac{g_{\p N \D}^2 \mu^2}{48 \p M^2 f_\p^2} 
\lc q_\D^3 + 3\, q_\D \, \frac{M^2 \sp m^2 \sm \m^2}{4 M^2} \lb (M \sp
m)^2 -\m^2 \rb \rc\;. 
\label{3.4}
\eeq

The values of the distance $R$ for which $\st_\D(R)/f_\p^2\m^2=1$ and
of the delta $\s$-term, calculated by means of eqs.(\ref{b.9},
\ref{b.10}), are given in table \ref{T3}, for different choices of the
coupling constants $g_{\p N\D}$ and $g_{\p \D\D}$. The $SU(4)$
predictions for these constants are 1.33 and 0.75, whereas the value
1.47 for the former yields the empirical decay width. The value 0.67
for the latter was used in ref.\cite{BHM}. Results for the real
component of $\s_\D$ are sensitive to the coupling constant $g_{\p \D
\D}$ and fully consistent with that given in ref.\cite{LGFD}, namely
$\s_\D=(32\pm 3)$ MeV. On the other hand, our prediction is larger than
that quoted in ref.\cite{BHM}. The values for the imaginary component
$\s_\D^I$, obtained by means of eq.(\ref{b.9}), are identical with
those given by eq.(\ref{3.4}), as they should.

\vspace{5mm}

\ni

\begin{table}[h]
\begin{center}
\caption{Real and imaginary parts of the $\D$ $\s$-term as function of
the coupling constants.} 
\vspace{3mm}
\begin{tabular} {|c|c|c|c|c|}
\hline 
$g_{\p N\D}$ & $g_{\p\D\D}$ & $R$ (fm) & Re $\s_\D$ (MeV) & Im $\s_\D$
(MeV) \\ \hline
\hline 
1.33         &  0.75        & 0.54     & 32.1	          &  -21.7 \\ \hline 
1.33         &  0.67        & 0.52     & 28.1	          &  -21.7 \\ \hline 
1.47  	     &  0.75	    & 0.55     & 31.7	          &  -26.7 \\ \hline 
1.47  	     &  0.67	    & 0.53     & 27.8	          &  -26.7 \\ \hline 
\end{tabular}
\label{T3}
\end{center}
\end{table}

\vspace{3mm}

\section{summary}
\label{4}

We have discussed a model aimed at determining $\s-$terms, which
consists in cutting off configuration space expressions at the point
where the cosine of the chiral angle becomes larger than 1. The model
has been used to calculate $\s_N$ and $\s_\D$ with success. In the
former case, a value very close to that extracted from experiment by
Gasser, Leutwyler and Sainio\cite{GLS}, was obtained. In the case of
the delta, the prediction 28 MeV $\leq \s_\D \leq$ 32 MeV, depending
on the coupling constants employed,  is also very close to the result
produced by another group\cite{LGFD}. The fact that the delta can
decay gives rise to a pion cloud which includes an oscillation and is
responsible for both the relation $\s_\D < \s_N$ and the consistency
of the imaginary part of $\s_\D$ with the decay width. Analytic
expressions also comply with chiral counting rules and give rise to
expected non-analytic terms in the chiral limit. These features
suggest that our calculational procedure is sound and can be reliably
applied to other systems.

\appendix
\section{loop integrals}
\label{A}

In the triangle diagrams, $p$ and $p'$ are the initial and final
baryon momenta, whereas $k$ and $k'$ are the momenta of the exchanged
pions. We also employ the variables  
\beq
q=(p \sm p') \;, \;\;\;\;\;\;\; P= (p\sp p')/2  \;, \;\;\;\;\;\;\; Q= (k\sp k')/2 \;.
\label{a.1}
\eeq

In all diagrams, the external baryon, with mass $m_e$, is assumed to
be on shell and one has  
\bea
p^2 =p'^2= m_e^2  \;, \;\;\; && \;\;\;\; P\cdot q=0\;,
\label{a.2}\\
\bar{u} \qs\, u = 0  \;, \;\;\; && \;\;\;\;  \bar{u} \Ps \, u = m \, \bar{u} \, u \;,
\label{a.3}\\
\bar{u}^\m  \qs \, u^\n = 0  \;, \;\;\; && \;\;\;\;  \bar{u}^\m \Ps \,
u^\n = M \, \bar{u}^\m \, u^\n \;. 
\label{a.4}
\eea

The basic loop integrals needed in this work involve either two or three denominators.
We use the definition 
\beq
\int [\cdots] =  \int \frac{d^4Q}{(2\p)^4} \; 
\frac{1}{[(Q \sp q/2)^2 \sm \m^2]\;[(Q\sm q/2)^2 \sm \m^2]}
\label{a.5}
\eeq

\ni
and the dimensionless expressions
\bea
&& I_{\p\p} =  \int [\cdots] 
=\frac{i}{(4\p)^2} \; \P_{\p\p}^{(00)}\;,
\label{a.6}\\[2mm]
&& I_{\p\p}^{\m\n} =  \int [\cdots] \frac{Q^\m Q^\n}{\m^2}
=\frac{i}{(4\p)^2} \; \lb g^{\m\n}\; \Pb_{\p\p}^{(00)}+ \cdots \rb \;,
\label{a.7}\\[2mm]
&& I_{x\p\p} =  \int [\cdots] \; \frac{2 \m m_e}{[(Q \sp P)^2 \sm m_x^2]}
=\frac{i}{(4\p)^2} \; \P_{x \p \p}^{(000)}\;,
\label{a.8}\\[2mm]
&& I_{x \p \p }^\m =  \int [\cdots]\; \frac{(Q^\m / \m)\;\;2 \m
m_e}{[(Q \sp P)^2 \sm m_x^2]} 
=\frac{i}{(4\p)^2}\; \lb \frac{P^\m}{m_e}\;\P_{x\p\p}^{(100)} + \cdots \rb \;,
\label{a.9}\\[2mm]
&& I_{x\p \p }^{\m\n} =  \int [\cdots]\; \frac{(Q^\m Q^\n / \m^2)\;\;2
\m m_e}{[(Q \sp P)^2 \sm m_x^2]} 
=\frac{i}{(4\p)^2}\; \lb g^{\m\n}\; \Pb_{x\p\p}^{(000)} + \cdots \rb \;,
\label{a.10}\\[2mm]
&& I_{x\p\p}^{\m\n\r} =  \int [\cdots] \; \frac{(Q^\m Q^\n Q^\r
/\m^3)\;\; 2 \m m_e}{[(Q \sp P)^2 \sm m_x^2]} 
=\frac{i}{(4\p)^2}\; \lb g^{\m\n} \frac{P^\r}{m_e} \;
\Pb_{x\p\p}^{(100)} +  \cdots \rb \;. 
\label{a.11}
\eea

\ni
where the ellipses indicate terms that do not contribute to the scalar
form factors.  
The usual Feynman techniques for loop integration allow one to write
the regular parts of these integrals as
\bea
&& \P_{\p\p}^{(00)} = - \int_0^1d a\; \ln \lp \frac{D_{\p\p}}{\m^2} \rp   \;,
\label{a.12}\\[2mm]
&& \Pb_{\p\p}^{(00)} = - \int_0^1d a\; \frac{D_{\p\p}}{2\,\m^2} \,\ln
\lp \frac{D_{\p\p}}{\m^2} \rp   \;, 
\label{a.13}\\[2mm]
&& \P_{x\p\p}^{(k00)} = - \int_0^1d a\;a \int_0^1 d b \;
[-m_e(1\!-\!a)/\m]^k \; \lp \frac{2 \m m_e}{D_{x\p\p}}\rp  \;, 
\label{a.14}\\[2mm]
&& \Pb_{x\p\p}^{(k00)} = - \frac{m_e}{\m}\;\int_0^1d a\;a \int_0^1 d b
\; [-m_e(1\!-\!a)/\m]^k \; 
\ln \lp \frac{D_{x\p\p}}{2 \m m_e}\rp  \;,
\label{a.15}
\eea

\ni
with
\bea
&& D_{\p\p} = \m^2 - a(1 \sm a)\;q^2\;,
\label{a.16}\\[2mm]
&& D_{x\p\p} = a \; \m^2 + (1 \sm a)\; m_x^2 - a\, (1\sm a) \; m_e^2
- a^2 \,b\,(1\sm b)\,q^2 \;. 
\label{a.17}
\eea

The dimensionless configuration space functions $S$ are defined  as
\beq
S = \int \frac{d\bk}{(2\p)^3}\; e^{-i\bk \cdot \bx}\; \P \;.
\label{a.18}
\eeq 

\ni
with $x=\m r$ and $\bk=\bq/\m$.
Performing the Fourier transforms, we find
\bea
S_{\p\p}^{(00)} &=&  \frac{1}{\p x^2} \; K_1(2x) \;,
\label{a.19}\\[2mm]
\Sb_{\p\p}^{(00)} &=& - \frac{1}{2\p x^4} \lb x\, K_0(2x) +K_1(2x)\rb  \;,
\label{a.20}\\[2mm]
\phi^2 >0 \rar S_{x\p\p}^{(k00)} &=& -\, \frac{2 m_e}{\m}\,\frac{1}{\p x} \,
\int_0^1 d a \; \frac{[-m_e(1\sm a)/\m]^k}{a} \; K_0 (2 \phi x) \;,
\label{a.21}\\[2mm]
\phi^2 < 0 \rar S_{x\p\p}^{(k00)} &=& \frac{m_e}{\m}\,\frac{1}{x} \,  
\int_0^1  d a \; \frac{[-m_e(1\sm a)/\m]^k}{a} \; \lb Y_0(2 |\phi| x) +
i  J_0(2 |\phi| x) \rb \; ,
\label{a.22}\\[2mm]
\phi^2 >0 \rar \Sb_{x\p\p}^{(k00)} &=& \frac{1}{\p x^2}\, \frac{m_e}{\m}\,  
\int_0^1 d a \;a [-m_e(1 \sm a)/\m]^k \; \phi \; K_1(2 \phi x) \;,
\label{a.23}\\[2mm]
\phi^2 < 0 \rar \Sb_{x\p\p}^{(k00)} &=& - \frac{1}{2 x^2}
\frac{m_e}{\m} \!\!   
\int_0^1 \!\! d a \;a [-m_e(1 \sm a)/\m]^k \;|\phi|\; \lb Y_1(2 |\phi| x) +
i  J_1(2 |\phi| x) \rb  , 
\label{a.24}
\eea

\ni
with
\beq
\phi^2 = [ a\,\m^2  + (1 \sm a)\, m_x^2 - a\,(1 \sm a)\, m_e^2] / (\m^2 \,a^2) \;.
\label{a.25}
\eeq

\section{scalar form factors}
\label{B}

We give here the expressions for $\tilde{\s}_{B_I}$,  due to triangle diagrams containing 
external states $B$ and an intermediate states $I$.  
The following interaction lagrangians\cite{D,H,HHK,BL} are used 
\bea 
&& \cL_{\p NN} = \frac{g_A}{2\, f_\p}\; \lc \bar{N}\, \g_\m \g_5
\,\tau_a \, N \rc \cdot \d^\m  \phi _a \;, 
\label{b.1}\\[2mm]
&& \cL_{\p N\D} = \frac{g_{\p N\D}}{f_\p}\; 
\lc \bar{\D}^\m \lb g_{\m\n}- (Z \sm 1/2) \g_\m \g_\n \rb M_a \, N \rc
\cdot \d^\n  \phi _a + h.c. \;, 
\label{b.2}\\[2mm]
&& \cL_{\p\D\D} = - \frac{g_{\p\D\D}}{f_\p} 
\lc \bar{\D}^\m \lp g_{\m\n}\g_\l - g_{\m\l} \g_\n - g_{\l\n} \g_\m
\rp \g_5 \, T_a \, \D^\n \rc \cdot  \d^\l \phi_a \;, 
\label{b.3}
\eea

\ni
where $\phi$, $N$ and $\D$ denote pion, nucleon and delta fields,
$f_\p$ is the pion decay constant, $g_A$, $g_{\p N\D}$ and
$g_{\p\D\D}$ are coupling constants, and $\tau$, $M$ and $T$ are
matrices that couple nucleons and deltas into isospin 1 states, with
$\tau_a \, \tau_a = 3$, $M_a^\dagger \, M_a = 2$, $M_a \, M_a^\dagger
=1$,  and $T_a^\dagger \, T_a =15/4$. For the coupling constants we
use $g_A=1.25$ and the $SU(4)$ results $g_{\p N\D}= 3 \,g_A /
2\sqrt{2}$ and $g_{\p\D\D} = 3\, g_A/5$. We also use $f_{\p}=93$~MeV,
$\m=139.57$~MeV, $m=938.27$~MeV and $M=1232$~MeV.

\vspace{1mm}
\newpage
\ni
{\bf nucleon:}
\vspace{1mm}

Using the loop integrals $S$ defined in appendix \ref{A}, we obtain the
following contributions to the nucleon scalar form factor 
\bea
&& \tilde{\s}_{N_N}(r) = \frac{3}{4} \lb \frac{\m \, g_A}{4\p\,f_\p}\rb^2  2m (\m^3)
\lc S_{\p\p}^{(00)}- S_{N\p\p}^{(100)} \rc \;,
\label{b.4}\\[4mm]
&& \tilde{\s}_{N_\D}(r) =  2 \lb \frac{\m \, g_{\p N
\D}}{4\p\,f_\p}\rb^2  \frac{(m \sp M)}{6 M^2}  (\m^3) 
\lc \sm \lb (m \sp M)\, (2M \sm m) \sp 2 \m^2 
\sp  \frac{m \,\m^2  (1 \sm  \bnb^2/2 )}{(m \sp M)} \rb S_{\p\p}^{(00)}\right.
\nn\\[2mm]
&& \left. - \frac{2 m \,\m^2}{(m \sp M)}\bar{S}_{\p\p}^{(00)}
\right.
\nn\\[2mm]
&& \left. + \frac{1}{2m\m} \lb (m^2 \sm M^2) \, (m \sp M)\, (2M \sm m) 
+ 2 \m^2 \, (m^2 \sm M^2) 
+ 6\, M^2\, \m^2 \, (1 \sm \bnb^2/2) \rb S_{\D\p\p}^{(000)} \right.
\nn\\[2mm]
&& \left. +  \frac{1}{2m} \,\lb (m \sp M) \, (4mM \sm M^2 \sm m^2) - 2 \m^2 (2M \sm m) 
+ \frac{6 M^2\,\m^2}{(m \sp M)} \,  (1\sm \bnb^2/2) \rb  S_{\D \p \p }^{(100)} \rc\;.
\label{b.5}
\eea

\vspace{1mm}
\ni
{\bf delta:}
\vspace{1mm}

In the evaluation of the triangle diagram, the external deltas are on
shell and one has the constraints $p\cd\;u^s (\bp)=\g\cd \;u^s (\bp)
=\ub^{s'}(\bp')\cd p'  = \ub^{s'}(\bp') \cd \g =0$. The $T$ matrix can
be cast in the form  
\beq
i T =  \ub_\m^{s'}(\bp') 
\lc \int \frac{d^4Q}{(2\pi)^4}\; \frac{\Theta^{\m\n}}{[ (Q\!-\!q/2)
^2\! -\!\m^2] [(Q\!+\!q/2)^2 \!-\!\m^2]} \rc  \;u_\n^s (\bp) \;, 
\label{b.6}
\eeq

\ni
with
\bea
&& \Theta_N^{\m\n}  = 2 \lb \frac{\m \, g_{\p N\D}}{f_\p}\rb^2 
(Q \sp q/2)^\m (Q \sm q/2)^\n \; \frac{m + M + \Qs}{\pb^2 \sm m^2}  \;,
\label{b.7}\\[2mm]
&& \Theta_\D^{\m\n}  = \frac{15}{4} \lb \frac{\m \, g_{\p \D\D}}{f_\p}\rb^2 
\lb g^{\m\n} \lp 2M - \frac{4M^2}{\pb^2 \sm M^2} \rp \Qs 
-\, \frac{8}{3} (Q \sp q/2)^\m (Q \sm q/2)^\n \,  \frac{M - \Qs}{\pb^2 \sm m^2}\rb \;,
\label{b.8}
\eea

\ni
with $\pb = P \sp Q$.
Performing the integrals, comparing the results with eq.(\ref{2.1}),
and going to configuration space, we obtain the contributions  
\bea
&& \tilde{\s}_{\D_N}(r) =  \lb \frac{\m\, g_{\p N\D}}{4\p\, f_\p}\rb^2 \m (\m^3)
\lb \frac{(m+M)}{2M} \, \Sb_{N\p\p}^{(000)} + \frac{\m}{2M}\, \Sb_{N\p\p}^{(100)}\rb  \;,
\label{b.9}\\[2mm]
&& \tilde{\s}_{\D_\D}(r) = \frac{15}{4} \lb \frac{\m \,
g_{\p\D\D}}{4\p\,f_\p}\rb^2  2M (\m^3) 
\lb S_{\p\p}^{(00)}-  S_{\D\p\p}^{(100)} 
- \frac{2\m}{3M}\;\Sb_{\D\p\p}^{(000)} + \frac{2 \m^2}{3M^2}\;\Sb_{\D\p\p}^{(100)} \rb\;.
\label{b.10}
\eea

\section{chiral symmetry}
\label{C}

In this appendix we show that results (\ref{b.4},\ref{b.5}) and
(\ref{b.9},\ref{b.10}) are fully compatible with standard chiral power
counting by means of a covariant chiral expansion\cite{HR}. It is
important to note that these expressions contain a factor $(\m^3)$,
which comes from the definition of the configuration space function
$S$ and must not be included in the counting. With this previous in
mind, we use the following relations among integrals, 	 
\bea
&& S_{\p\p}^{(00)} = \lb 1 - \frac{\m^2 \bnb^2}{4 m_e^2} \rb \, S_{x\p\p}^{(100)}
+\lb \frac{\m}{2 m_e}\, (1-\bnb^2/2) + \frac{(m_e^2 \sm m_x^2)}{2 \m
m_e}\rb  \, S_{x\p\p}^{(000)} \;, 
\label{c.1}\\[2mm]
&& 2 \lb 1 - \frac{\m^2 \bnb^2}{4 m_e^2}\rb \, \Sb_{x\p\p}^{(000)}
=  \lb \frac{\m}{2 m_e}\, (1 \sm \bnb^2/2) + \frac{(m_e^2 \sm
m_x^2)}{2 \m m_e}\rb  \, S_{\p\p}^{(00)} 
\nn\\[2mm]
&& \;\;\;\;\;\; - \lb \frac{(m_e^2 \sm m_x^2)^2}{4 \m^2 m_e^2} -
\frac{m_e^2 \sp m_x^2}{2 m_e^2} 
+ \frac{\m^2}{4 m_e^2} + \frac{m_x^2}{4 m_e^2} \,\bnb^2  \rb \, S_{x\p\p}^{(000)} \;,
\label{c.2}\\[2mm]
&& 2 \lb 1 - \frac{\m^2 \bnb^2}{4 m_e^2}\rb \, \Sb_{x\p\p}^{(100)}
= - \frac{1}{3} (1-\bnb^2/4) \, S_{\p\p}^{(00)}
\nn\\[2mm]
&& \;\;\;\;\;\; - \lb \frac{(m_e^2 \sm m_x^2)^2}{4 \m^2 m_e^2} -
\frac{m_e^2 \sp m_x^2}{2 m_e^2} 
+ \frac{\m^2}{4 m_e^2} + \frac{m_x^2}{4 m_e^2} \,\bnb^2  \rb \, S_{x\p\p}^{(100)} \;,
\label{c.3}
\eea

\ni
which are obtained by multiplying eqs.(\ref{a.8}-\ref{a.10}) by
$P_\m$, neglecting short range terms with a single pion propagator,
and going to configuration space. In the case $m_e=M$ and $m_x=m$, the
expansion of eqs.(\ref{a.8}-\ref{a.9}) yield the leading order
relations 
\bea
&& \P_{\D\p\p}^{(000)} \simeq \frac{(2\,m_e\,\m)}{(m^2 \sm M^2)}\;\P_{\p\p}^{(00)}\;,
\label{c.4}\\[2mm]
&& \P_{\D\p\p}^{(100)} \simeq -\, \frac{(2\,m_e\,\m)^2}{(m^2 \sm
M^2)^2}\; \Pb_{\p\p}^{(00)}\;. 
\label{c.5}
\eea

Truncating the expansions at $\cO(q^4)$, we find
\bea
&& \tilde{\s}_{N_N}(r) = \frac{3}{4} \lb \frac{\m \, g_A}{4\p f_\p}\rb^2 \m\, (\m^3)
\lb (1 \sm \bnb^2/2) \,S_{N\p\p}^{(000)}- \frac{\m}{2m}\bnb^2 S_{\p\p}^{(00)} \rb \;,
\label{c.6}\\[4mm]
&& \tilde{\s}_{N_\D}(r) = -\, \frac{4}{3} \lb \frac{\m \, g_{\p N
\D}}{4\p\,f_\p}\rb^2  \frac{\m^2}{(M \sm m)} (\m^3) 
\lb (1 \sm \bnb^2/2) \, - \frac{m^2}{3 M^2} \, (1 \sm \bnb^2/4) \rb S_{\p\p}^{(00)} \; ,
\label{c.7}\\[4mm]
&& \tilde{\s}_{\D_N}(r) = \frac{1}{4} \lb \frac{\m\, g_{\p N\D}}{4\p\,
f_\p} \rb^2 \frac{(M \sp m)}{M} \m (\m^3) \lb \frac{(M \sp m)}{4\m M^2}
(M^2 \sm m^2) S_{N\p\p}^{(100)} \right . 
\nn \\ [2mm]
&& + \left . \lp 1 - \frac{M \sm m}{6M} \rp \lp 1-\frac{\bnb ^2}{4}
\rp S_{N\p\p}^{(000)} \rb \;,
\label{c.8}\\[2mm]
&& \tilde{\s}_{\D_\D}(r) = \frac{15}{4} \lb \frac{\m \,
g_{\p\D\D}}{4\p\,f_\p}\rb^2 \m\, (\m^3) \,  
\lb (1 \sm \bnb^2/2)\, S_{\D\p\p}^{(000)}- \frac{\m \,\bnb^2}{2M}
S_{\D\p\p}^{(100)} \right. 
\nn\\[2mm]
&& - \left. \frac{4}{3}\;\Sb_{\D\p\p}^{(000)} + \frac{4
\m}{3M}\;\Sb_{\D\p\p}^{(100)} \rb\;. 
\label{c.9}
\eea

These results, except for eq. \ref{c.8}, which contains
imaginary terms, are compatible with chiral counting
rules. Contributions begin at $\cO(q^3)$ for diagrams in which
internal and external baryons are identical and at $\cO(q^4)$ when this
does not happen.

\section{chiral limit}
\label{D}

In this section we show that our model for the nucleon $\s$-term is
consistent with the standard ChPT expansion. In the paper by Becher and
Leutwyler\cite{BL}, one finds, using our notation 
\bea
&& \s_N = -4c_1\,\m^2 - \frac{9\,g_A^2\,\m^3}{64\p f_\p^2}
-\frac{3 \m^4}{16\p^2 f_\p^2 m}\lp g_A^2 \sm 8 c_1m \sp c_2m \sp 4c_3m
\rp \,\ln \frac{\m}{m} 
\nn\\[2mm]
&& - \frac{3 \m^4}{64\p^2 f_\p^2 m}\lp 3 g_A^2 \sm 8 c_1m \sp 4c_3m \rp 
+ 2 \bar{e}_1\;,
\label{d.1}
\eea

\ni
where $c_i$ and $e_1$ are, respectively, low enegy constants (LECs)
from the $\cL_N^{(2)}$ and $\cL_N^{(4)}$ lagrangians. The bar over
$e_1$ indicates that it has been renormalized.

In order to expand our $\s_N$, we use in eq.(\ref{c.4}) the result 
\beq
S_{N\p\p}^{(000)} \simeq - \frac{e^{-2x}}{2\, x^2}+
\frac{\m}{m\p\,x^2}\lb x K_0(2x) + K_1(2x) \rb \;, 
\label{d.2}
\eeq

\ni
which holds\cite{HRR} for $\m/m << 1$.
This allows integrations in eq.(\ref{2.9}) to be performed
analytically and one finds 
\bea
&& \s_N = \frac{4}{3} \p R^3 \;f_\p^2 \m^2
+ \frac{3\,g_A^2 \m^3}{16\p f_\p^2}\lc \lp \frac{1}{4} \sp \frac{1}{2\m R}\rp e^{-2\m R}
- \frac{\m}{2m\p}\lb 4 K_0(2\m R) \right.\right.
\label{d.3}\\[2mm]
&& \left.\left.
\sp \lp 2\m R \sp \frac{6}{2\m R}\rp K_1(2\m R) \rb \rc 
+ \frac{g_{\p N\D}^2 \m^4}{6\p^2 f_\p^2 (M\sm m)}
\lb K_0(2\m R) \sp \lp 3 \sm  \frac{m^2}{2 M^2}\rp \frac{K_1(2\m R)}{2\m R} \rb\;.
\nn
\eea

An  expansion for small values of $\m$ yields
\bea
&& \s_N = \frac{4}{3} \p R^3 \;f_\p^2 \m^2
\label{d.4}\\[2mm]
&&
+ \frac{3\,g_A^2}{16\p f_\p^2}\lb \frac{\m^2}{2 R^2} \lp R \sm \frac{1}{m \p}\rp 
- \frac{3\, \m^3}{4}
-\frac{\m^4}{m\p}\ln \m R + \frac{\m^4\, (5 \sm 4\g)}{4m\p} 
+ \frac{\m^4 \, R}{2} \rb 
\nn\\[2mm]
&&
+ \frac{g_{\p N\D}^2}{12 \p^2 f_\p^2 (M\sm m)}
\lb \frac{\m^2}{R^2} \lp 3  \sm \frac{m^2}{2M^2} \rp 
+ \m^4 \lp 4 - \frac{m^2}{M^2}\rp \ln \m R
+ \m^4 \lp \sm 3 \sp 4\g + \frac{m^2}{2M^2} (1 \sm 2\g) \rp \rb\;,
\nn
\eea

\ni
where $\g$ is the Euler constant. This result reproduces the first
three terms of eq.(\ref{d.1}),  provided one absorbs the factors
proportional to $\m^2$ into the definition of $c_1$, uses the delta
contributions to the $c_i$, which are given by  
\beq
c_1^\D = 0\;, \;\;\;\;\; 
c_2^\D = \frac{4\,g_{\p N\D}^2 m^2}{9 M^2 (M\sm m)}\;, \;\;\;\;\; 
c_3^\D = \frac{- 4\, g_{\p N\D}^2}{9 (M\sm m)}\;,
\label{d.7}
\eeq

\ni and chooses the value $R=1/m$ for the cutting  radius. As the
renormalized constant $\bar{e}_1$ contains factors proportional to
$\m^4$, terms of this kind need not to coincide. 

\begin{acknowledgments}
The works by I.P.C. and D.O.S. were supported by CNPq and the work by
G.R.S.Z., by FAPESP (Brazilian agencies).
\end{acknowledgments}


\end{document}